\documentclass[
reprint,
amsmath,amssymb,
aps,
prr,
longbibliography,
floatfix
]{revtex4-2}

\usepackage{graphicx}% Include figure files
\usepackage{color}
\usepackage[english]{babel}
\usepackage[utf8]{inputenc}
\usepackage{dcolumn}% Align table columns on decimal point
\usepackage{bm}% bold math
\usepackage{hyperref}
\usepackage[capitalise]{cleveref}
\usepackage[mathlines]{lineno}% Enable numbering of text and display math
\usepackage{multirow}
\usepackage{braket}
\def\ket#1{| #1 \rangle}
\def\bra#1{\langle #1|}

\newcommand{\Det}{\Delta \omega }

\usepackage{MnSymbol}
\usepackage{physics}

\begin{document}

\preprint{APS/123-QED}

\title{Harnessing Time Symmetry to Fundamentally Alter Entanglement in Photoionization}

\author{Axel Stenquist
}
\author{Jan Marcus Dahlström$^{\text{\dagger}}$
}

\affiliation{Department of Physics, Lund University, 22100 Lund, Sweden. }

\begin{abstract}
\noindent 
The Grobe--Eberly doublet phenomenon occurs in photoelectron distributions when a field dresses the remaining ion. Its manifestation is due to entanglement between a free electron and a hybrid state of light and matter. Direct detection of such entanglement is however not possible by coincidence schemes due to the dressing mechanism having an inconspicuous phase correlation effect on the ion. Here, it is shown that odd envelopes fundamentally alter the entanglement, such that channel-resolved photoelectron distributions become identifiable in coincidence with the internal state of the field-free ion. This constitutes a first usage of the parity of time symmetry in strong-field interactions. 
\end{abstract}
  
\maketitle

\begin{table}[b!]
\begin{flushleft}
  $^\text{\dagger}$ marcus.dahlstrom@matfys.lth.se
\end{flushleft}
\end{table}

\section{Introduction}

\begin{figure}[t!]
    \centering
    \includegraphics[width = 0.5\textwidth]{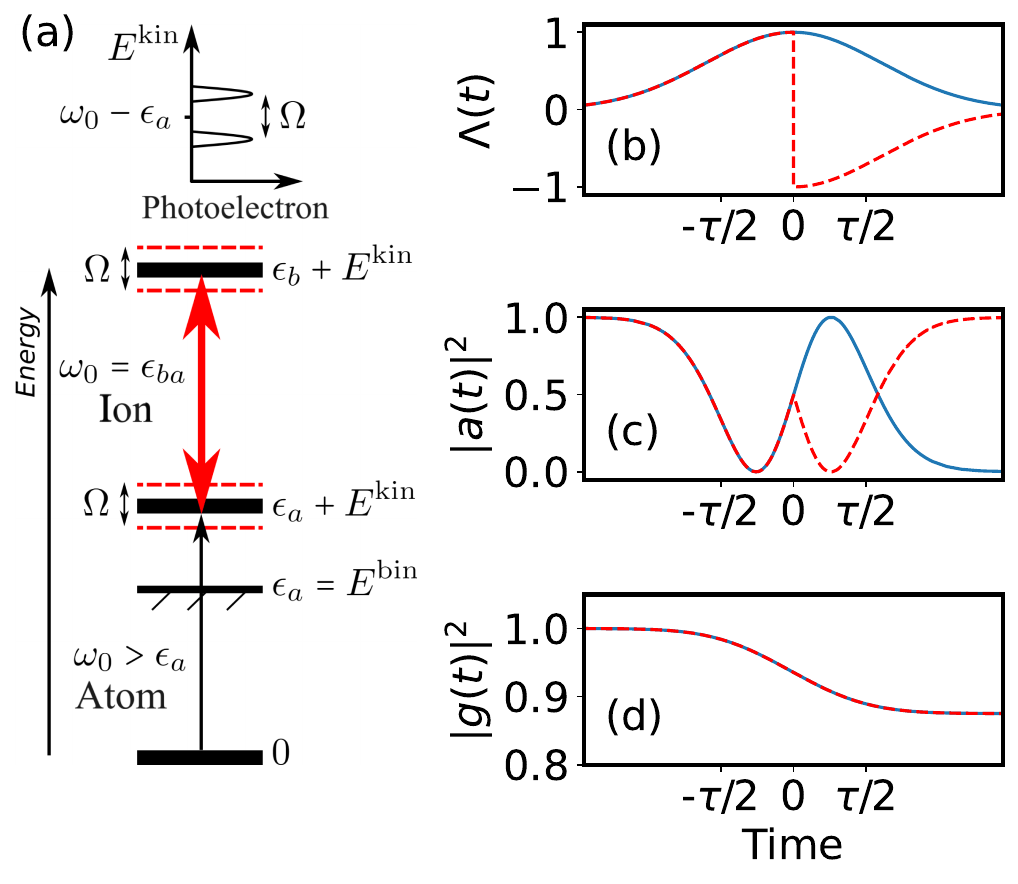}
\caption{\textit{Grobe--Eberly doublet and odd envelopes.} (a) Schematic illustrating the Grobe--Eberly doublet formation due to atomic photoionization and sequential Rabi coupling in the ion. (b) Pulse envelope for a Gaussian, $\Lambda_\tau^\text{even}(t)$, (blue) and zero-Gaussian pulse, $\Lambda_\tau^\text{zero}(t)$, (red dashed). (c) Corresponding ground state population for an atom with one-photon Rabi coupling to an excited state. (d) Same as (c), but with one-photon coupling to continuum via photoionization.}
\label{fig1}
\end{figure}

Symmetry is indispensable for understanding the interaction of particles in quantum physics. It provides rules for angular momenta in atoms, band structure in crystals, and governs the gauge freedom of fundamental fields \cite{cohentannoudji_photons_1997}. 
Conservation laws, derived from Noether's theorem have also been studied for open quantum systems \cite{albert_symmetries_2014}. 
It can be considered a resource for control \cite{manzano_harnessing_2018}, or as a benchmark for studies of symmetry-broken processes \cite{even_tzur_selection_2021,tzur_selection_2022}. 
Interactions between particles and coherent light pulses are intrinsically time- and space-dependent, but symmetry provides the ground for such quantum transitions, where photoionization exemplifies an open-system process \cite{rohringer_strongly_2012,bostrom_time-stretched_2018}.   
Quantum entanglement and decoherence between free particles, created through the photoelectric effect, have been studied using synchrotron radiation with coincidence detection \cite{akoury_simplest_2007,piancastelli_new_2017}, and more recently with time-resolution provided by attosecond pulses \cite{nishi_entanglement_2019,vrakking_control_2021,busto_probing_2022,koll_experimental_2022}.
The development of free-electron lasers (FEL) enables production of high-intensity ultra-short pulses in the extreme-ultraviolet (XUV) and X-ray regime that are utilized in a multitude of unique strong-field investigations  \cite{young_femtosecond_2010,lindroth_challenges_2019,young_roadmap_2018,rudenko_femtosecond_2017,kanter_unveiling_2011}. Seeded FELs, such as FERMI \cite{allaria_highly_2012}, additionally provide temporal coherence, allowing for coherent control experiments  \cite{fushitani_femtosecond_2016,prince_coherent_2016,maroju_attosecond_2023}. In particular, seeded FELs allow for strong coupling at XUV wavelengths \cite{nandi_observation_2022}, which is essential for quantum operations on ultrafast timescales. Recently an investigation of quantum entanglement in the photoelectric effect, generated between photoelectrons and light-dressed atomic ions, was conducted at FERMI \cite{nandi_generation_2024}. In agreement with the seminal theoretical work of Grobe and Eberly \cite{grobe_observation_1993}, and more recent investigations \cite{walker_observation_1995,hanson_manifestations_1997,zhang_photoemission_2014,yu_core-resonant_2018}, a doublet was observed in the photoelectron spectrum 
corresponding to two dressed states of the ion, as illustrated in \cref{fig1}(a). 
While the physical mechanism for this effect was attributed to quantum entanglement \cite{nandi_generation_2024},  its degree could not be measured. The theory of entanglement in ultrafast light-driven processes is a rapidly expanding subject 
\cite{fedorov_packet_2004,you_attosecond_2016,majorosi_quantum_2017,bostrom_time-stretched_2018,ruberti_quantum_2022,maxwell_entanglement_2022,ishikawa_control_2023,ruberti_bell_2024,stammer_quantum_2023,laurell_continuous-variable_2022}, but the role of explicit time symmetries in such interactions has not been studied.

In this article, we propose that quantum entanglement between photoelectrons and ions in the strong-coupling regime can be fundamentally altered by harnessing the time symmetry of the interaction between the coherent light field and the atom. We predict that: i) the generation of entanglement can be delayed by a factor of two, and ii) the photoelectrons from different ionic states will ``avoid each other''. The latter point is an essential step forward in the field, as it opens up for coincidence detection of entanglement in the strong-coupling regime. 
Atomic units are used throughout this text, $e=\hbar=m_e=4\pi\epsilon_0=1$, unless otherwise stated.

\section{Theory}

The dynamics of an atom subjected to photoionization with subsequent dressing of the ion was first studied by Grobe and Eberly \cite{grobe_observation_1993}. Using Laplace transformation, they showed that it is possible to find simple analytical solutions for flattop envelopes. The decay of the atom then follows a simple exponential law, as expected for a bound state coupled to a continuum \cite{cohentannoudji_atomphoton_1998}, while the photoelectron distribution develops into an Autler--Townes-like doublet with two peaks separated by the Rabi frequency. In our previous work, we have revisited this phenomenon using a quantum optics formalism, see the supplemental material of Ref.~\cite{nandi_generation_2024}, and described how quantum entanglement is created between the photoelectron and the dressed ion. In this work, we are interested in more complex interactions with time-dependent envelopes. To this end, we turn to a semi-classical analytical model, similar to the one proposed by Yu and Madsen \cite{yu_core-resonant_2018}, where we apply explicit symmetries of even and odd envelopes. Using time-inversion symmetries, we obtain simple expressions for the photoelectron distributions that depend on a single time argument, and only a single set of ``Rabi amplitudes''. This makes systematic exploration of quantum entanglement feasible.  

Electron dynamics are computed within the dipole approximation, driven by a time-dependent classical electric field, 
\begin{equation}
E(t) = E_0\Lambda_\tau(t) \sin(\omega_0 t),
\end{equation}
where $E_0$, $\Lambda_\tau(t)$, $\omega_0$ and $\tau$ are the amplitude, envelope, central frequency and pulse length of the field, respectively. 
In resonant interactions with two-level atoms, the state amplitudes:  $a(t) = \cos[\theta(t)/2]$ and $b(t) = \sin[\theta(t)/2]$, follow the pulse area: 
\begin{equation}
\theta(t) = \int^t_{-\infty} dt' \Omega(t'),
\end{equation}
which is the angle that the Bloch vector rotates from the ground state \cite{mccall_self-induced_1967,feynman_geometrical_1957}. Here, 
\begin{equation}
\Omega(t) = z_{ba} E_0 \Lambda_\tau(t) \in \mathbb{R},
\end{equation}
is the Rabi frequency. It is useful to also define the {\it absolute} pulse area for the total envelope: $\theta_\tau= \int_{-\infty}^{\infty}dt \abs{\Omega(t)}$. The pulse area should not be confused with the integrated electric field area \cite{plachenov_pulses_2023}.

\begin{figure*}[t!]
    \centering
    \includegraphics[width = 1\textwidth]{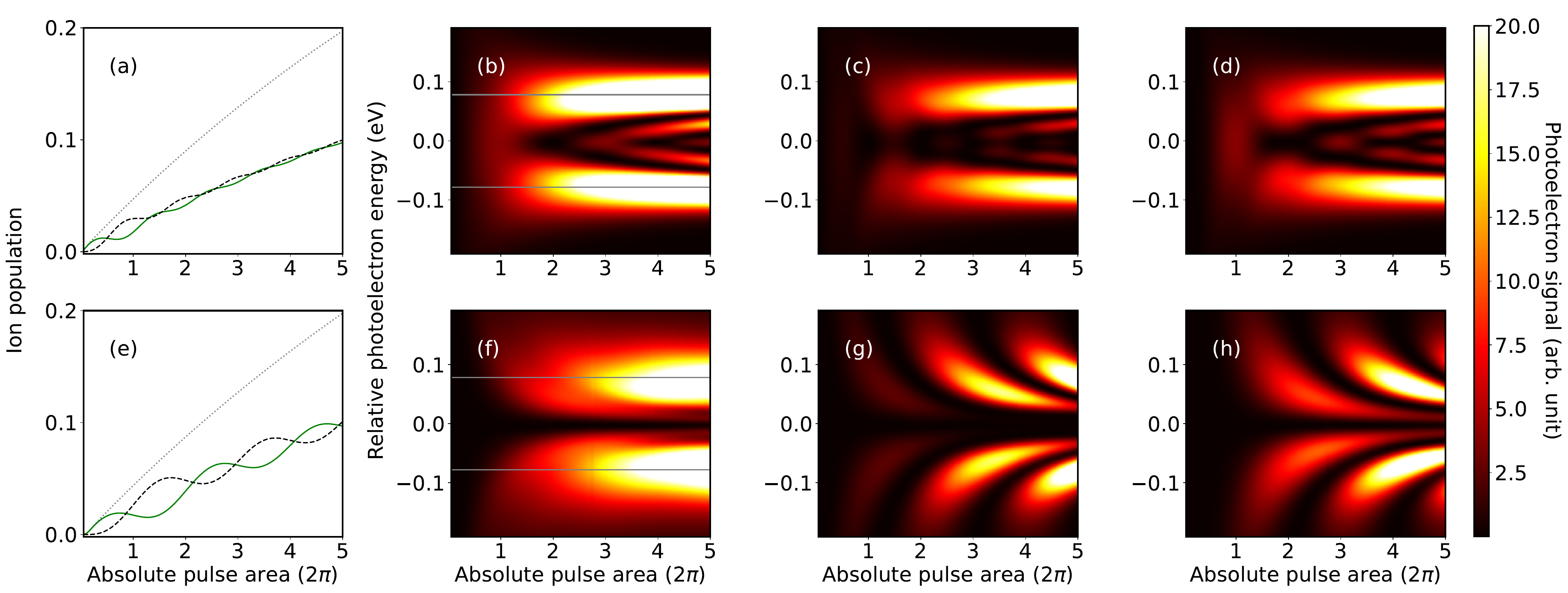}
\caption{\textit{Build-up of the ion dynamics and photoelectron spectra over absolute pulse area.} \textbf{Top row:} results for Gaussian pulses. (a) shows the ion populations of the ground state (green), the excited state (black dashed) and the full ionic population (grey dotted). The photoelectron probability distribution
is presented in (b) and the ion-channel-resolved photoelectron spectra corresponding to the ground state and excited state channels are presented in (c) and (d), respectively. \textbf{Bottom row:} corresponding results for zero-Gaussian pulses. Heat maps are saturated at high photoelectron signals. Constant peak intensity and resonant frequency are used (see main text).}
\label{fig2}
\end{figure*}

\subsection{Harnessing Time Symmetry}
% \textit{Harnessing time symmetry---}
Envelopes used to describe pulses in experiments are often positive even functions, $\Lambda_\tau^\mathrm{even}(t)=\Lambda_\tau^\mathrm{even}(-t)\ge 0$, such as Gaussian or flattop pulses. While modest deformations of even envelopes constitute a form of symmetry breaking, the other parity --  the odd envelopes -- could, from a symmetry point of view, generate dramatically different physics. Here, we consider odd pulses generated from even envelope functions:  
\begin{equation} \label{Eq:zero_envelope}
\begin{split} 
\Lambda_\tau^\mathrm{zero}(t) 
&= \lim_{K\rightarrow \infty} -\frac{2}{\pi}\atan(Kt)\Lambda_\tau^\mathrm{even}(t)  \\
&= -\text{sign}(t) \Lambda_\tau^\mathrm{even}(t).
\end{split}
\end{equation}
In particular, we will study the Gaussian and its corresponding zero-area envelope, the ``zero Gaussian'', but our conclusions generally apply to other pulse envelopes \cite{vasilev_complete_2006}. Consistent results for smooth sign changes are presented in \cref{sec_Smooth} below. In practice, odd effective envelopes can be constructed using delayed Gaussian pulses \cite{ardini_generation_2024} or by dichromatic fields \cite{he_coherently_2019}.
As is well-known from quantum optics, zero-area pulses have a vanishing area at the end of the pulse --- a property trivially fulfilled for odd envelopes: 
$\lim_{t\rightarrow\infty}\theta(t)=0$. A curious feature of zero-area pulses is that they induce transient dynamics in a two-level system, but eventually return the system to its initial state \cite{rhodes_influence_1968,grieneisen_observation_1972}. Over the last decades, reshaping of pulses to zero area through atomic/molecular medium has moved from picosecond \cite{rothenberg_observation_1984} to femtosecond timescales \cite{matusovsky_0_1996}. Recently, the generation of zero-area pulses in perfect transient absorption with attosecond pulses was proposed \cite{he_resonant_2022}. The odd envelopes considered here are special symmetric cases of zero-area pulses, which have so far been overlooked in the strong-field community. 

\subsection{Resonant Two-Level Dynamics}
% \textit{Resonant two-level dynamics---}
A comparison of the $3\pi$-area Gaussian envelope and its corresponding zero-area envelope is presented in \cref{fig1}(b). The Gaussian pulse causes Rabi flopping to the excited state, while the zero-area pulse induces a time reversal of the dynamics at the centre of the pulse, bringing the population back to the ground state, as shown in \cref{fig1}(c). Clearly, the atomic populations are even functions for odd envelope pulses. The state amplitudes are also even functions, as is evident from their area dependence. We stress that state amplitudes from even envelopes are not symmetric in time (except for $2n\pi$ area pulses, where $n$ is an integer).

\subsection{Photoionization Dynamics}
% \textit{Photoionization dynamics---}
The photoelectric effect predicts that the kinetic energy of the photoelectron is $E^\text{kin} = \omega_0 - E^\text{bin}$, where $E^\text{bin}$ is the binding energy of the atom. In contrast to the two-level dynamics, the photoelectric effect is Markovian, in the sense that the ionization rate is unaffected by the sudden sign change of the envelope. The dipole coupling from the atomic ground state, $g$, to the ionic ground state, $a$: $\Omega_0^{ag}=z_{ag}E_0$, is assumed to be independent of the photoelectron energy, $E^\text{kin}$. In \cref{fig1}(d), we show that the survival probability of the atomic ground state, driven by Gaussian and zero-Gaussian pulses, are indeed identical. (However, the photoelectron wavepacket is changed.)

\subsection{Grobe--Eberly Doublet}
% \textit{Grobe--Eberly doublet---} 
The combination of photoionization with subsequent resonant two-level dynamics, was first studied by Grobe and Eberly \cite{grobe_observation_1993}. A modification to the photoelectric effect was predicted: the formation of doublet peaks separated by the Rabi frequency \cite{grobe_observation_1993}. Given the relative photoelectron energy, $\epsilon = E^\text{kin} - (\omega_0 - E^\text{bin})$, the photoelectron peaks appear at the energies $\epsilon_\pm = \pm \overline{\Omega}/2$, corresponding to two final dynamically dressed ionic states.

The dynamics can be well explained using a 2-step model \cite{yu_core-resonant_2018}: {\it --1--} The field ionizes the atom at time $t$, 
leaving the ion in the ground state and creating a photoelectron with the remaining energy. {\it --2--} The field induces Rabi oscillations in the ion, between its ground and excited state with detuning $\Delta\omega=\omega_0-(\epsilon_b-\epsilon_a)$,
while the photoelectron propagates freely.  
Here, we apply time symmetries (TS) to find the following simple expressions for the final state amplitudes $\alpha_\tau(\epsilon)$ and $\beta_\tau(\epsilon)$ corresponding to the ground $\ket{a,\epsilon}$ and excited $\ket{b,\epsilon}$ ionic states, respectively, with a photoelectron of energy $\epsilon$: 
\begin{equation} \label{Eq: full amp}
    \begin{split}
        \alpha_\tau(\epsilon) &= \frac{\Omega^{ag}_0}{2} \!\!\! 
        \int_{-\infty}^\infty \!\!\!\! dt \,  a^{(\text{TS})}_\tau( -t) 
        \Lambda^{\! (\text{TS})}_\tau(t) g_{\tau}( t) e^{i \epsilon t}  , \\
        \beta_\tau(\epsilon) &=  \frac{\chi \Omega^{ag}_0}{2} \!\!\! 
        \int_{-\infty}^\infty \!\!\!\! dt \,  {b^{(\text{TS})}_\tau}^*\!\! ( -t) 
        \Lambda^{\! (\text{TS})}_\tau(t) g_{\tau}( t) e^{i (\epsilon-\Det) t}  ,
    \end{split}
\end{equation}
where $a_\tau^{(\text{TS})}(t)$, $b_\tau^{(\text{TS})}(t)$ and $g_\tau(t)$ are ionic ground state, the ionic excited state and the atomic ground state amplitudes, respectively. The state amplitudes are expressed within the interaction picture and computed for the explicit time symmetries of the envelope (TS: being \textit{even} or \textit{odd}) with pulse duration, $\tau$. The ionic state amplitudes in \cref{Eq: full amp} depend only on a single time argument that describes propagation from its ground state, $a$, at $-\infty$ to the intermediate time $-t$. In this way, state amplitudes can be efficiently computed for envelopes that satisfy the TS requirement, where $\chi$ is 1 (-1) for even (odd) symmetry, see \cref{sec_A} for more details. We note that \cref{Eq: full amp} is valid for both resonant and non-resonant cases.

\section{Results}

We consider the experimentally relevant helium photoionization process, where the field is resonant with the ionic transition between 1s and 2p. The peak intensity of the Gaussian and zero-Gaussian envelopes is set to $I_0 = 1.25 \cdot 10^{13}$ W/cm$^2$, corresponding to a maximum Rabi frequency of $\Omega_0 = z_{ba}E_0 \approx 0.2$ eV, and an average Rabi frequency of $\overline{ \Omega }\approx 0.16$ eV, as defined in \cref{sec_A}. Pulse durations, $\tau$, up to 81 fs are considered, corresponding to absolute pulse areas $0 \leq \theta_\tau \leq 10\pi$. 
First, we present the ion dynamics $\int \abs{\alpha_\tau(\epsilon)}^2 d\epsilon$ and $\int \abs{\beta_\tau(\epsilon)}^2 d\epsilon$, followed by the photoelectron spectrum $\abs{\alpha_\tau(\epsilon)}^2 + \abs{\beta_\tau(\epsilon)}^2$, where $\abs{\alpha_\tau(\epsilon)}^2$ and $\abs{\beta_\tau(\epsilon)}^2$ are the ion-channel resolved photoelectron probability distributions for the ground and excited state, respectively.

\subsection{Ion Dynamics}
% \textit{Ion dynamics---} 
For a Gaussian pulse, we observe oscillations in the ionic state populations as the pulse area is increased, as presented in \cref{fig2}(a). 
The oscillations are attenuated with increasing pulse area, presumably due to the range of instantaneous intensities. The number of oscillations roughly aligns with the pulse area (here up to five modulations) in good agreement with prior predictions \cite{zhang_photoemission_2014,yu_core-resonant_2018}. In contrast, we find that for the zero-Gaussian pulse, in  \cref{fig2}(e), the ion goes through half the number of oscillations, 2.5 modulations, for the absolute area of $10\pi$, with less attenuation.

\subsection{Photoelectron Spectra}
% \textit{Photoelectron spectra---} 
In \cref{fig2}(b), we present the pulse-area-resolved final-state photoelectron spectra for Gaussian pulses. The spectra develop from a single peak, centred on the resonant photoelectron energy, $\epsilon = 0$, to the dynamical Grobe--Eberly doublet, $\pm \overline{\Omega}/2$ (grey lines). Spectral fringes between the peaks, related to the number of completed Rabi cycles for single pulses, are observed in agreement with prior predictions for Gaussian pulses \cite{zhang_photoemission_2014}. This phenomenon is found also in ordinary two-photon resonant ionization, where the photon couples to the resonance before ionization \cite{olofsson_photoelectron_2023}. It is known that such substructures are similar for flattop and smooth envelopes (for sufficiently small area pulses) \cite{rogus_resonant_1986,simonovic_manifestations_2023}. 
The channel-resolved photoelectron spectra are presented in (c) and (d) for the ionic ground and excited state, respectively. The doublet forms earlier in the ground state ($\theta \approx 2\pi$), whereas in the excited state it forms later ($\theta \approx 3\pi$) \cite{zhang_photoemission_2014,yu_core-resonant_2018}. The channel-resolved substructures are oscillatory, (c) and (d), but ``piece together'' to form continuous fringes in the combined signal (b), similar to quantum interference effects predicted for ordinary resonant two-photon ionization \cite{olofsson_photoelectron_2023}.

In \cref{fig2}(f), we show that zero-Gaussian pulses yield photoelectron distributions that differ from the dynamical Grobe--Eberly doublet phenomenon in three ways. Firstly, no peak is formed at the energy $\epsilon = 0$, 
which means that the usual photoelectric pathway is ``blocked''. Secondly, the photoelectron spectra for larger areas form a doublet with significantly wider peaks. Thirdly, the spectral fringes between the doublet peaks \cite{zhang_photoemission_2014} are missing in the odd envelope case. 
The difference between Gaussian and zero-Gaussian pulses is more striking in the channel-resolved photoelectron distributions. 
The two channels do not overlap, in other words: they ``avoid each other'', as observed in (g) and (h) corresponding to the ground and excited state, respectively. 
While related spectral phenomena have been observed in pump-probe schemes~\cite{wollenhaupt_control_2003,yu_core-resonant_2018}, the underlying role of time symmetry has not been discussed.  

\begin{figure}
    \centering
    \includegraphics[width=0.99\linewidth]{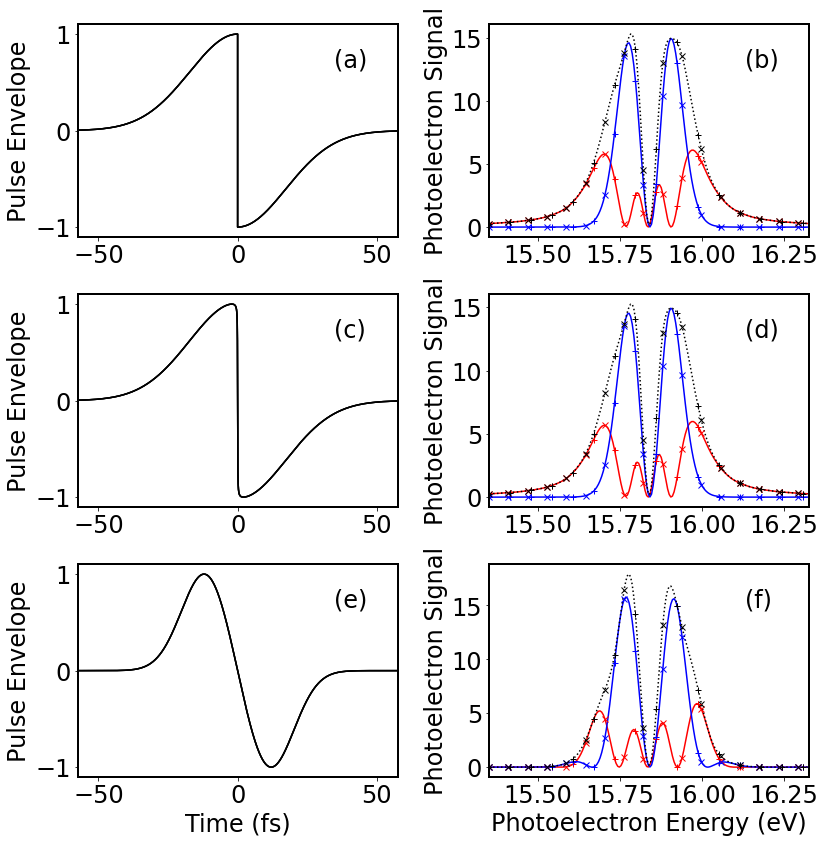}
    \caption{\textit{Photoelectron spectra for odd envelopes with varying shape and CEP.} The photoelectron spectra of the odd pulse (a), smooth odd pulse (c) and double Gaussian pulse (e) are presented in (b), (d) and (f) respectively, where $\abs{\alpha(\epsilon)}^2$ is shown in red, $\abs{\beta(\epsilon)}^2$ in blue and $\abs{\alpha(\epsilon)}^2+\abs{\beta(\epsilon)}^2$ in dotted black lines. CEP values of $\phi=0$,  $\phi=\frac{\pi}{4}$ and $\phi=\frac{\pi}{2}$ are represented by lines, crosses and plus signs, respectively.}
    \label{fig:smooth_envelope}
\end{figure}
\section{Discussion}

Here, our use of odd envelope symmetry implies a {\it phase discontinuity} of the driving field \cite{wollenhaupt_quantum_2006}, leading to distinct and useful behaviour of the photoelectrons, as will be discussed below.
While smooth even envelopes form dynamically dressed states, 
the dynamics from odd envelopes is different due to the associated phase discontinuity. Thus, the smooth evolution of dressed states is prevented and electron distributions with novel properties are produced. 

For the zero-Gaussian pulse, the central peak does not form at $\epsilon = 0$. This is clear from \cref{Eq: full amp} since the depletion is small, $g_\tau(t)\approx 1$, the envelope, $\Lambda_\tau^\text{zero}(t)$, is odd and the state amplitudes, $a_\tau(t)$ and $b_\tau(t)$, are even for all pulse durations. Hence, the integral is zero, inhibiting resonant photoionization. 

\subsection{Zero-Flattop Model}

An analytical expression for the final state amplitudes can be determined in the resonant case ($\Delta\omega=0$) for the ``zero flattop'' envelope by neglecting depletion, as described in \cref{sec_B}, and can be written on the form $c_j = c_j^+ + c_j^-$ as,  
\begin{equation}\label{Eq: analytical final main}
\begin{split}
\alpha_\tau^\pm(\epsilon) &= \pm \frac{i\Omega^{ag}_0}{2} 
\frac{\cos(\Omega_0 \tau/4)-\cos( \epsilon \tau/2)}{\Omega_0/2 \mp \epsilon},  
\\
\beta_\tau^\pm(\epsilon) &= \pm \frac{i\Omega^{ag}_0}{2} 
\frac{\sin(\Omega_0 \tau/4) \mp \sin( \epsilon \tau/2)}{\Omega_0/2 \mp \epsilon}, 
\end{split}
\end{equation}
where the origin of the doublet peaks is observed in the denominator, centred on $\pm\Omega_0/2$. We note that the photoelectron densities of the two channels avoid each other, due to their cosine and sine character for the ionic ground and excited state, respectively. 
By comparing with the analytical model for the even flattop pulse, see \cref{Eq: analytical even}, we note that the even pulse yields oscillations at twice the frequency of the zero-area pulse, as observed in \cref{fig2}~(a) and (e), respectively.% and \cref{fig3}. 

\subsection{Phase Discontinuous Smooth Pulses}\label{sec_Smooth}
% \textit{Smooth pulses---}
Throughout this work, odd envelopes, have been defined through the simple relation \cref{Eq:zero_envelope}. Such a pulse, with $4\pi$ absolute pulse area, and its corresponding photoelectron spectra are presented in the top row of \cref{fig:smooth_envelope}, in (a) and (b), respectively. This is compared to an envelope with a smooth sign change:
\begin{equation}\label{Eq:odd pulse smooth}
    \Lambda_\tau^\mathrm{zero}(t) = - \frac{2}{\pi}\atan(K t) \Lambda_\tau^\mathrm{even}(t),
\end{equation} 
in the middle row for $K=1/2$. 
%This envelope is equivalent to \cref{Eq:zero_envelope} in the limit $K \rightarrow \infty$. 
Additionally, results are presented in the bottom row for a pulse composed of two Gaussians, shifted in time, one with a positive and the other a negative envelope. This is equivalent to the two Gaussians having a carrier-envelope phase (CEP) difference of $\pi$ (carrier-phase discontinuity). We note that the photoelectron spectra attained for these three pulses are qualitatively similar. Hence, the results are not sensitive to the discontinuity of the derivative of the envelope. The results are insensitive to the CEP value (as denoted by the lines, crosses and plus signs).

\subsection{General Stationary Phase Analysis}

A more general analytical expression, valid for any reasonable {\it odd} pulse envelope, is given by
\begin{equation}\label{Eq: analyt general}
\begin{split}
\alpha(\epsilon) &= -i\eta(\epsilon)\sin[\varphi(\epsilon)],  
\\
\beta(\epsilon) &= 
- i\eta(\epsilon)\cos[\varphi(\epsilon)],  
\end{split}
\end{equation}
where the real functions $\eta(\epsilon)$ and $\varphi(\epsilon)$ are found through Eq.~(\ref{Eq: full amp}) for the resonant case ($\Delta\omega=0$), using the stationary phase approximation, as described in \cref{sec_B2}. Remarkably, Eq.~(\ref{Eq: analyt general}) is a general result for odd envelopes having any odd number of $\pi$-phase discontinuities. The analysis yields pairs of stationary solutions that explain the observed avoiding behavior in \cref{fig2}(g) and (h). Mathematically, in each pair of stationary solutions, one comes from the positive Euler term and the other from the negative Euler term of the ionic interaction amplitudes. Thus, the different characters ($\cos$ or $\sin$) of the ionic interaction amplitudes is transferred to the channel-resolved photoelectron distributions making them distinguishable. 

We mention that general {\it even} envelopes, which may contain an even number of $\pi$-phase discontinuities, do not satisfy Eq.~(\ref{Eq: analyt general}). Using the stationary phase approximation, pairs of stationary solutions are found, but in contrast to the odd case, they both come from the {\it same} Euler term, which leads to a similar interference structure ($\cos$) in the photoelectron distributions for both ionic channels, in agreement with our numerical results in \cref{fig2}(c) and (d). Thus, it is time symmetry -- not phase discontinuity -- that is at the heart of the avoiding phenomenon.

\begin{figure}[t!]
    \centering
    \includegraphics[width = 0.5\textwidth]{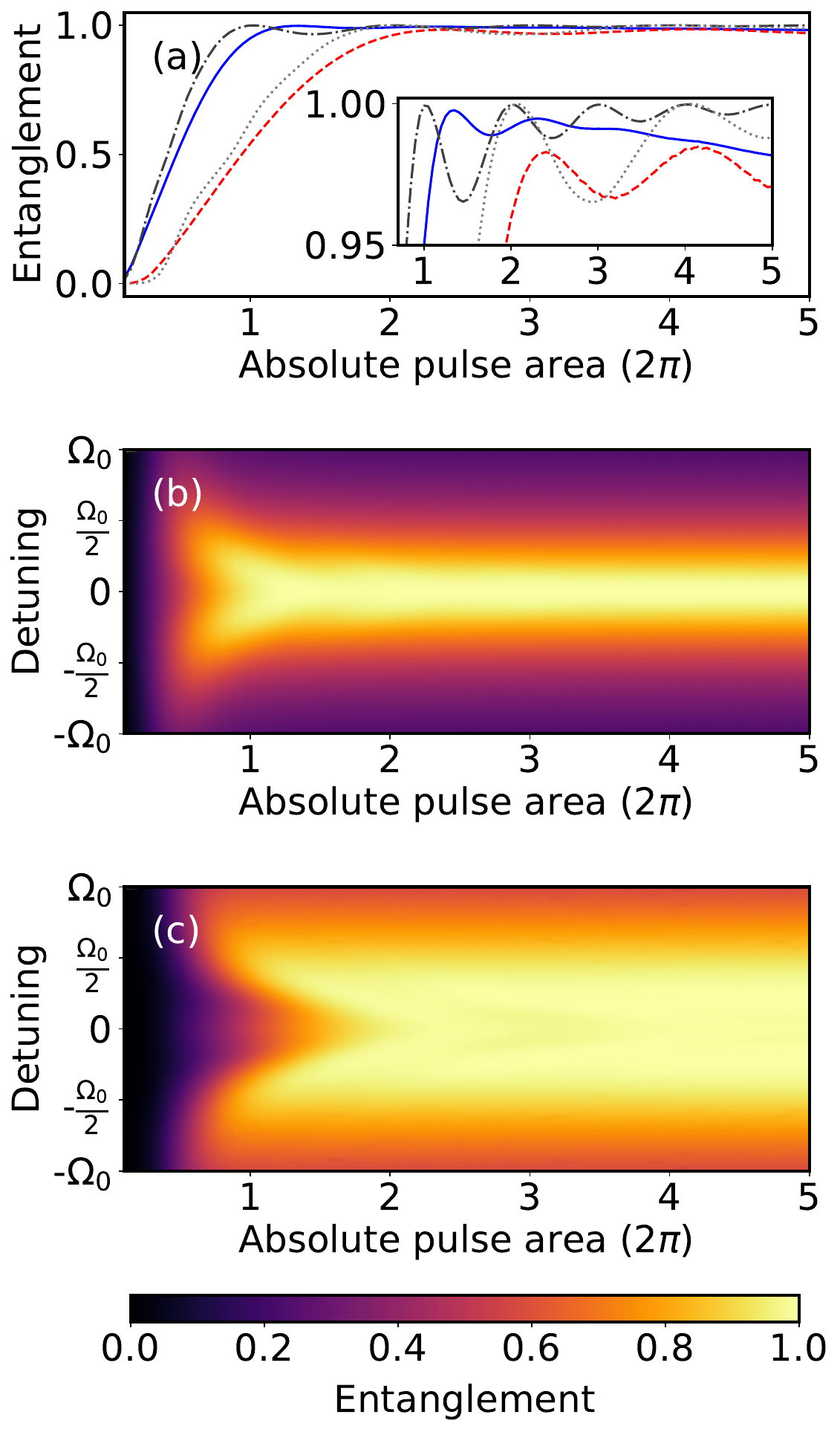}
\caption{\textit{Entanglement resolved over absolute pulse area and detuning.} Absolute-pulse-area resolved entanglement (a) for a resonant Gaussian (blue) and its corresponding zero-area pulse (red dashed) as well as a flattop (dark grey dash-dotted) and flattop zero-area pulse (light grey dotted). Detuning- and pulse-area-resolved entanglement is presented in (b) and (c) for the Gaussian and zero-Gaussian pulses, respectively.}
\label{fig3}
\end{figure}

\subsection{Entanglement}
% \textit{Entanglement---} 
The degree of entanglement is quantified using the von Neumann entropy of entanglement 
\begin{equation}
S_\text{vN}(\tau) = - \Tr{\rho^{I}(\tau) \log_2[\rho^{I}(\tau)]},
\end{equation}
where $\rho^{I}(\tau)$ is the $2 \times 2$ reduced post-measurement density matrix of the ion \cite{haroche_exploring_2013}, see \cref{sec_C} for details. The entanglement, resolved over pulse area, is presented in \cref{fig3}(a) for Gaussian and zero-Gaussian pulses. Additionally, the flattop-envelope entanglement is given. The insert shows the entanglement between 0.95 and 1. Compared to the flattop pulses, we see a slight delay in the build-up of entanglement for smooth pulses. 
In contrast, the entanglement is severely delayed in the case of zero-area pulses, where the flattop becomes fully entangled at $\theta = 4\pi$. The zero-area Gaussian is further slightly delayed reaching a peak entanglement of $\sim$98\%. 

% \subsection{Detuning}
% \textit{Detuning---} 
By introducing a detuning to the pulse, we can control the build-up of the entanglement. Detuning- and absolute-pulse-area-resolved entanglement is shown in \cref{fig3}(b) and (c) for the Gaussian and its corresponding zero-area pulse, respectively. In both cases, the entanglement decreases as the pulse is detuned, but the build-up of entanglement becomes faster. Such behaviour is consistent with the associated weaker, but faster, Rabi oscillations for detuned interactions.

\section{Conclusions}

In this article, we have studied the entanglement between photoelectrons and strongly coupled ions. Our results show that odd envelopes, a special class of zero-area pulses, break the dressing mechanism of the ion and fundamentally change the channel-resolved photoelectron distributions. The resonant photoelectric effect is blocked and entanglement is delayed by a factor of two in terms of absolute pulse area. It is found that the entanglement still reaches values close to those maximally allowed for qubit systems $S_\text{vN}\approx1$, but its manifestation is changed: the electrons now avoid each other when detected in coincidence with the internal states of the field-free ion [as shown in \cref{fig2}(g,h)]. 
We propose that experimental discrimination of the excited ionic state can be performed by photoionization using a subsequent low-frequency laser field: $\gamma+A^{+*}\rightarrow A^{2+}+e^-$, whereas the tightly bound ground state would be unaffected. 
Thus, our work provides key insights for future strong-coupling entanglement experiments, bringing the usage of zero-area pulses from single-photon physics and quantum information applications  \cite{derouault_one-photon_2012,costanzo_zero-area_2016,he_coherently_2019,lipka_single-photon_2021}, to the strong-field physics of electrons and ions. Our results also provide insights into how {\it temporal decoherence} of light fields (modelled by phase discontinuities) can reduce or delay the establishment of quantum entanglement in photoionization. 

\begin{acknowledgments}
We acknowledge discussions with Edvin Olofsson, Yijie Liao, and Saikat Nandi. JMD acknowledges support from the Olle Engkvist Foundation: 194-0734, the Knut and Alice Wallenberg Foundation: 2019.0154, and the Swedish Research Council: 2024-04247.
\end{acknowledgments}

\appendix

\section{Theoretical Methodology} \label{sec_A}

% The dynamics of an atom subjected to photoionization with subsequent dressing of the ion was first studied by Grobe and Eberly \cite{grobe_observation_1993}. Using Laplace transformation, they showed that it is possible to find simple analytical solutions for flattop envelopes. The decay of the atom then follows a simple exponential law, as expected for a bound state coupled to a continuum \cite{cohentannoudji_atomphoton_1998}, while the photoelectron distribution develops into an Autler--Townes-like doublet with two peaks separated by the Rabi frequency. In our previous work, we have revisited this phenomenon using a quantum optics formalism, see the supplemental material of Ref.~\cite{nandi_generation_2024}, and described how quantum entanglement is created between the photoelectron and the dressed ion. In this work, we are interested in more complex interactions with time-dependent envelopes. To this end, we turn to a semi-classical analytical model, similar to the one proposed by Yu and Madsen \cite{yu_core-resonant_2018}, where we apply explicit symmetries of even and odd envelopes. Using time-inversion symmetries, we obtain simple expressions for the photoelectron distributions that depend on a single time argument, and only a single set of ``Rabi amplitudes''. This makes systematic exploration of quantum entanglement feasible.  

The atom (Helium) is initially in the atomic ground state, $\ket{g}$ (1s$^2$), with energy $0$. Photoionization by a field takes the atom into a state $\ket{a,E^\text{kin}}$, containing the ionic ground state, denoted $a$ (1s), with energy $\epsilon_a = E^\text{bin}$, and the  photoelectron with energy $E^\text{kin}$. The field induces Rabi oscillations between the ground state and the excited ionic state, denoted $b$ (2p) with energy $\epsilon_b$. The energy separating the ionic states is $\epsilon_b - \epsilon_a = 40.8$ eV. The corresponding excited composite state is $\ket{b,E^\text{kin}}$. The interaction with the field is described within the dipole approximation as 
\begin{equation}\label{Eq: interaction}
    V_\tau(t) = \hat z E_0 \Lambda_\tau(t) \sin(\omega_0 t),
\end{equation}
where $\hat z$ is the dipole operator, $E_0$ is the field amplitude, $1\ge\Lambda_\tau(t)\ge-1$ is the normalized field envelope and the central frequency is $\omega_0 = \epsilon_b - \epsilon_a + \Det$, where $\Det$ is the detuning. The pulse duration, $\tau$, describes the full width at half max of the intensity profile. The Rabi frequency, described as $\Omega(t) = z_{ba} E_0 \Lambda_\tau(t) \in \mathbb{R}$, couples the ground state $\ket{a}$ to the excited state $\ket{b}$, through the dipole matrix element, $z_{ba}=0.373$. We define the average Rabi frequency over a smooth envelope as 
\begin{equation}
    \overline{\Omega} = \int dt\abs{\Omega(t)}^3 / \int dt \abs{\Omega(t)}^2.
\end{equation}
To compute the dynamics, we calculate the evolution of the wave function inspired by the resolvent-operator method. The time dependence of the envelope is incorporated as a slowly varying coupling to the continuum, while explicit time-dependent interaction with the photoelectron--ion state is considered. There is no coupling back to the neutral system from the photoelectron--ion states, as is the case for the dynamics in the complement space for the resolvent operator technique \cite{cohentannoudji_atomphoton_1998}. For simplicity, the photoelectron is assumed to be unaffected by the laser field after ionization, and its angular momentum is therefore fixed as a $p$-wave. By inserting unity between the operators, and simplifying the expression using the selection rules, we attain a final wave function, 
\begin{equation} \label{Eq: Dyson WF}
\begin{split}
    \ket{\Psi(\tau)} \approx  &  \;
    \ket{g}\bra{g}U_0(\infty,-\infty)\ket{g} \\ 
    &- i \int dt \int dE^\text{kin} \, U(\infty,t) \ket{a,E^\text{kin}} \\
    &\times \bra{a,E^\text{kin}} V_\tau(t) \ket{g}\bra{g} U_0(t,-\infty)\ket{g},
\end{split}
\end{equation}
where our handling of the time-dependent depletion of the atom turns out be to equivalent to that proposed by Yu and Madsen using a flat continuum (dynamical Stark-shifts are neglected) \cite{yu_core-resonant_2018}, yielding
\begin{equation} \label{Eq: depletion}
    \bra{g} U_0(t,-\infty)\ket{g} = \exp[- \frac{\pi z_{ag}^2 E_0^2}{4}\int_{-\infty}^t dt \Lambda_\tau^2(t)]= g_\tau(t),
\end{equation}
where $z_{ag}\approx 0.502$ is the dipole matrix element between the atomic ground state of helium and the continuum, which is assumed to be constant around the energy $E^\text{kin} = \omega_0 - E^\text{bin}$. The interaction between the atomic and ionic ground states is described as 
\begin{equation} \label{Eq: ag interaction}
    \bra{a,E^\text{kin}} V_\tau(t) \ket{g} = \Omega_0^{ag}\Lambda_\tau(t)\sin(\omega_0 t).
\end{equation}
As described in our previous work \cite{nandi_generation_2024}, the evolution operator $U(\infty,t) = U_e(\infty,t) U_I(\infty,t)$ is factorized into the propagator of the free propagation of the photoelectron 
\begin{equation} \label{Eq: photoelectron prop}
    U_e(\infty,t)\ket{a,E^\text{kin}} = e^{-iE^\text{kin}(\infty-t)}\ket{a,E^\text{kin}},
\end{equation}
and the time evolution of the ion, which induces Rabi oscillations between the ground and excited ionic states, $a \leftrightarrow b$, described by the amplitudes $a_\tau(\infty,t)$ and $b_\tau(\infty,t)$. Thus, the ionic evolution in the quantum-optics interaction frame \cite{nandi_generation_2024,yu_core-resonant_2018}, 
\begin{equation}\label{Eq: interaction amplitude}
    \begin{split}
        U_I(\infty,t)\ket{a,E^\text{kin}} = 
        &a_\tau(\infty,t)\ket{a,E^\text{kin}}e^{-i \epsilon_a (\infty-t)}  \\
        + &b_\tau(\infty,t)\ket{b,E^\text{kin}}e^{-i (\epsilon_b - \omega_0) (\infty-t)},
    \end{split}
\end{equation}
determines the final state, at $\infty$, given that the propagation starts at time $t$. Using the symmetry of the envelope, we find that this can be expressed by the time-reversed Rabi oscillations. For a resonant pulse, the dynamics are determined by the pulse area, $\theta(\infty,t) = \Omega_0 \int_t^\infty dt' \Lambda_\tau(t')$. For an even pulse, this is equivalent to $\theta(-t,-\infty) = \Omega_0 \int^{-t}_{-\infty} dt' \Lambda_\tau(t')$. Thus, $a_\tau(\infty,t) = \cos[\theta(\infty,t)/2] = \cos[\theta(-t,-\infty)/2] = a_\tau(-t,-\infty)$, which we denote $a_\tau(-t)$, and similarly  $b_\tau(\infty,t) = b_\tau(-t)$. In contrast, for an odd-resonant pulse, $\theta(\infty,t) = - \theta(-t,-\infty)$. Because the ground state oscillation is even (of cos character), whilst the excited state is odd (of sine character) the amplitudes can be expressed as  $a_\tau(\infty,t) = a_\tau(-t)$ and $b_\tau(\infty,t) = -b_\tau(-t)$. 
In the detuned case, we find that the imaginary, and real parts of $b$ change sign for even and odd envelopes respectively. Hence, the state amplitudes for the even envelope can be expressed as $a_\tau(\infty,t) = a_\tau(-t)$ and $b_\tau(\infty,t) = b_\tau^*(-t)$, whilst for odd envelopes $a_\tau(\infty,t) = a_\tau(-t)$ and $b_\tau(\infty,t) = -b_\tau^*(-t)$.

Here $a_\tau(t)$ and $b_\tau(t)$ are the Rabi oscillating state amplitudes, computed numerically using a two-level essential-state system, driven by a field of arbitrary shape and detuning \cref{Eq: interaction}, from the initial state $\abs{a_\tau(-\infty)}^2=1$ and $\abs{b_\tau(-\infty)}^2=0$. To generalize our expression, we define the symmetry factor, $\chi = 1$ for even envelopes and $\chi=-1$ for odd. The final state coefficients, $\alpha_\tau(E^\text{kin})$ and $\beta_\tau(E^\text{kin})$ corresponding to the ground $\ket{a,E^\text{kin}}$ and excited $\ket{a,E^\text{kin}}$ ionic state, respectively, with a photoelectron of energy $E^\text{kin}$, are attained by projecting \cref{Eq: Dyson WF} on the ionic states, propagated to time $t = \infty$, and inserting \cref{Eq: depletion,Eq: ag interaction,Eq: photoelectron prop,Eq: interaction amplitude},
and defining the relative photoelectron energy $\epsilon = E^\text{kin} - \omega_0 + \epsilon_a$ we arrive at the expression for the complex amplitude of the final states, \cref{Eq: analytical final main},
% \begin{equation} \label{Eq: full amp_append}
%     \begin{split}
%         \alpha_\tau(\epsilon) &= \frac{\Omega_0^{ag}}{2} \!\!\int dt a_\tau^\text{(TS)}(-t)
%         \Lambda_\tau^\text{(TS)}(t) g_\tau(t)  
%         e^{i \epsilon t},
% \\
%         \beta_\tau(\epsilon) &= \chi \frac{\Omega_0^{ag}}{2} \!\!\int dt {b_\tau^\text{(TS)}}^*(-t)
%         \Lambda_\tau^\text{(TS)}(t) g_\tau(t) 
%         e^{i (\epsilon - \Det) t},
%     \end{split}
% \end{equation}
where we have added the superscript (TS) to denote the quantity's dependence on the time symmetry of the envelope.  

The full density matrix can then be constructed as, 
\begin{equation}
    \begin{split}
        \rho(\tau) = 
          &   \rho_{g\,g}(\tau)\ket{g}\bra{g} \\
        + &    \sum_j \int_0^\infty d\epsilon \left[ \rho_{g\,j} (\tau;\epsilon)\ket{g}\bra{j,\epsilon} 
            + \text{h.c.} \right] \\ 
        % + & \sum_j \int_0^\infty  d\epsilon \rho_{j \,g}(\tau;\epsilon)\ket{j,\epsilon}\bra{g} \\
        + & \sum_{jj'} \int_0^\infty d\epsilon \int_0^\infty d\epsilon' \rho_{j \,j'} (\tau;\epsilon,\epsilon')\ket{j,\epsilon}\bra{j',\epsilon'},        
    \end{split}
\end{equation}
where the density matrix elements are $\rho_{g \,g}(\tau) = g_\tau^*(\infty)g_\tau(\infty)$, $\rho_{g \,j}(\tau;\epsilon) = g_\tau^*(\infty) c_{j}(\tau,\epsilon)$ and $\rho_{j \,j'}(\tau;\epsilon,\epsilon') = c_j^*(\tau,\epsilon)c_{j'}(\tau,\epsilon')$, where $j\in\{a,b\}$ and $c_a(\tau,\epsilon) = \alpha_\tau(\epsilon)$ and $c_b(\tau,\epsilon) = \beta_\tau(\epsilon)$.

\section{Analytical Zero-Flattop Model}\label{sec_B}

An analytical model for the even-envelope flattop pulse is presented in \cite{nandi_generation_2024}. Here an analytical model is constructed for the case of the zero-area pulse, by neglecting depletion, $g_\tau(t) = 1$, and using a resonant odd flattop pulse, with envelope $\Lambda_\tau(t) = \text{sign}(t)[\Theta(t+\tau/2) - \Theta(t-\tau/2)]$, where $\Theta$ is the Heaviside function. Thus, the state coefficients are
\begin{equation} \label{Eq: analytical 1}
\begin{split}
        \alpha_\tau(\epsilon) 
        &= \frac{\Omega_0^{ag}}{2}   \int_{-\tau/2}^{\tau/2} dt   \;a_\tau(-t)\text{sign}(t) e^{i \epsilon t}, \\
        \beta_\tau(\epsilon) 
        &= -\frac{\Omega_0^{ag}}{2}   \int_{-\tau/2}^{\tau/2} dt   \;b_\tau(-t)\text{sign}(t) e^{i \epsilon t}, 
\end{split}
\end{equation}
Since $\text{sign}(t)$ is odd and both $a_\tau(-t) = a_\tau(t)$ and $b_\tau(-t) = b_\tau(t)$ are even for the zero-area pulse, the coefficients will be non-zero when $e^{i \epsilon t}$ is odd. Replacing $e^{i \epsilon t}\rightarrow i \sin(\epsilon t)$ makes the integrand even. Hence, $\int_{-\tau/2}^{\tau/2} dt \rightarrow 2\int_{-\tau/2}^{0} dt$. In the interval $t \in [-\tau/2, 0]$ we can insert the well-known Rabi amplitudes for the flattop pulse $a_\tau(t) = \cos(\Omega_0(t+\tau/2)/2)$ and $b_\tau(t) = \sin(\Omega_0(t+\tau/2)/2)$, where the oscillations are initialized at the start of the pulse $-\tau/2$. Thus, we can simplify the coefficients further as 
\begin{equation} \label{Eq: analytical 2}
\begin{split}
        \alpha_\tau(\epsilon) 
        &= i \Omega_0^{ag}   \int_{-\tau/2}^{0} dt \cos[\Omega_0(t+\tau/2)/2] \sin(\epsilon t),
        \\
        \beta_\tau(\epsilon) 
        &= -i \Omega_0^{ag}   \int_{-\tau/2}^{0} dt \sin[\Omega_0(t+\tau/2)/2] \sin(\epsilon t). 
\end{split}
\end{equation}
The integrand can be rewritten using the trigonometric product-to-sum identities, yielding, on the simplified form $c_j = c_j^+ + c_j^-$,  
\begin{equation} \label{Eq: analytical 3}
\begin{split}
        \alpha_\tau^\pm(\epsilon)
        &=\pm \frac{i \Omega_0^{ag}}{2}  \int_{-\tau/2}^{0} dt  
         \sin(\Omega_0 t/2 + \Omega_0 \tau / 4 \pm \epsilon t), 
        \\
        \beta_\tau^\pm(\epsilon)
        &=\pm  \frac{i \Omega_0^{ag}}{2}  \int_{-\tau/2}^{0} dt  
        \cos(\Omega_0 t/2 + \Omega_0 \tau / 4 \pm \epsilon t). 
\end{split}
\end{equation}
Integration yields the analytical coefficients \cref{Eq: analytical final main},
% \begin{equation}\label{Eq: analytical final}
% \begin{split}
% \alpha_\tau^\pm(\epsilon) &= \pm \frac{i\Omega^{ag}_0}{2} 
% \frac{\cos(\Omega_0 \tau/4)-\cos( \epsilon \tau/2)}{\Omega_0/2 \mp \epsilon},  
% \\
% \beta_\tau^\pm(\epsilon) &= \pm \frac{i\Omega^{ag}_0}{2} 
% \frac{\sin( \epsilon \tau/2) \mp \sin(\Omega_0 \tau/4)}{\Omega_0/2 \mp \epsilon}, 
% \end{split}
% \end{equation}
which have weight close to $ \epsilon = \pm \Omega_0/2 $ due to the denominator. Since the numerators are oscillating (over pulse duration $\tau = \theta/\Omega_0$ and photoelectron energy $\epsilon$), multiple peaks may emerge in the photoelectron distributions. As the oscillations in the ground state, $a$, are of cosine character and the oscillations in the excited state, $b$, are of sine character, the photoelectron peaks of the two ionic channels ``avoid each other''. This opens up for studies of entanglement by coincidence measurements (kinetic energy of photoelectron {\it vs.} internal state of the ion). 

We can compare this odd-envelope model with the analytical model for the flattop pulse presented in \cite{nandi_generation_2024} by writing the even envelope model on the same form (and neglecting depletion),
\begin{equation}\label{Eq: analytical even}
\begin{split}
\alpha_\tau^\pm(\epsilon) &= \pm \frac{i\Omega^{ag}_0}{4} 
\frac{e^{i \epsilon \tau} - e^{\pm i \Omega_0 \tau / 2}}{\Omega_0/2 \mp \epsilon},  
\\
\beta_\tau^\pm(\epsilon) &= \frac{\Omega^{ag}_0}{4} 
\frac{e^{i \epsilon \tau} - e^{\pm i \Omega_0 \tau / 2}}{\Omega_0/2 \mp \epsilon}.
\end{split}
\end{equation}
Here we note that the even envelope dynamics oscillate at twice the frequency, $\epsilon$ and $\Omega_0/2$, compared to the corresponding zero-area pulse case, which oscillated at $\epsilon/2$ and $\Omega_0/4$. Additionally, for the even-envelope case, the coefficients of the ground and excited states are of both cosine and sine character. Therefore, the ion-channel resolved photoelectron spectra do not show the oscillatory behaviour of the zero-area pulse case. The two ion channels therefore overlap, requiring measurements of the phase to observe entanglement.

\section{Stationary Phase Approximation}\label{sec_B2}

A more general analytic expression for the complex amplitudes is arrived at through the stationary phase approximation. In the resonant case the Rabi amplitude of the ground state is given through the area theorem as, $a(t) = \cos(\theta(t)/2)$, again depletion of the atom is neglected $g(t)=1$. The analytical expression \cref{Eq: full amp}, can then be simplified through partial integration as 
\begin{equation}
\begin{split}
    \alpha(\epsilon)
    &= - i\epsilon \frac{\Omega_0}{\Omega_0^{ag}}
    \int_{-\infty}^\infty \!\!\!\!dt \sin (\theta(t)/2)\exp(i\epsilon t) \\
    &= - i\epsilon \frac{z_{ba}}{2 i z_{ag}}
    \int_{-\infty}^\infty \!\!\!\!dt \left\{e^{i[\epsilon t+\theta(t)/2]} - e^{i[\epsilon t-\theta(t)/2]}\right\}.
\end{split}
\end{equation}
The stationary phase approximation can then be applied to the two terms in the integral, where the stationary points are given by $\epsilon + \Omega\Lambda(t_s)/2 = 0$ for the first Euler term and $\epsilon + \Omega\Lambda(t_r)/2 = 0$ for the second Euler term. By harnessing the symmetry of the odd envelope we find that $t_s=-t_r$. Thus, the amplitude can be described by the analytical expression
% \begin{equation}
%     \alpha_\tau(\epsilon) = - i\epsilon \frac{z_{ba}}{z_{ag}} \sqrt{\frac{4\pi}{\abs{\Omega_0 \Lambda'(t_s)}}} \sin\left[\epsilon t_s + \frac{\theta(t_s)}{2} + \kappa \frac{\pi}{4}\right]
% \end{equation}
\begin{equation}\label{Eq:amp_SPA_a}
    \alpha(\epsilon) =  \sum_{t_s}\frac{- i\epsilon z_{ba}2\sqrt{\pi}}{z_{ag}\sqrt{\abs{\Omega_0 \Lambda'(t_s)}}}  \sin\left[\epsilon t_s + \frac{\theta(t_s)}{2} + \kappa \frac{\pi}{4}\right]
\end{equation}
where $\Lambda'(t)$ denotes the derivative of the envelope and $\kappa = \text{sign}[\Omega_0\Lambda'(t_s)]$. By following the same procedure the excited state amplitude is determined to be 
\begin{equation}\label{Eq:amp_SPA_b}
    \beta(\epsilon) =  \sum_{t_s}\frac{- i\epsilon z_{ba}2\sqrt{\pi}}{z_{ag}\sqrt{\abs{\Omega_0 \Lambda'(t_s)}}}  \cos\left[\epsilon t_s + \frac{\theta(t_s)}{2} + \kappa \frac{\pi}{4}\right].
\end{equation}
Thus, as the amplitudes \cref{Eq:amp_SPA_a,Eq:amp_SPA_b} are composed of sums of sine and cosine functions, respectively, the amplitudes can be written as \cref{Eq: analyt general}. Hence, the avoiding behaviour of the photoelectron distributions of the two channels can be seen in the sine and cosine character of the two amplitudes. 

\section{Quantifying Entanglement}\label{sec_C}

In order to analyse our results, we form the post-measurement density matrix by conditioning the full density matrix on the photoionization, by applying the projection operator $Q = \int d\epsilon \sum_j \ket{j,\epsilon}\bra{j,\epsilon}$ and renormalizing with the population of ionization, $\Tr{\rho Q}$. The post-measurement density matrix is then $\tilde\rho = Q\rho Q / \Tr{\rho Q} = \sum_{jj'} \iint d\epsilon \, d\epsilon' \rho'_{j\, j'}(\tau;\epsilon,\epsilon') \ket{j,\epsilon}\bra{j',\epsilon'}$. The degree of entanglement is quantified using the von Neumann entropy of entanglement $S_\text{vN}(\tau) = - \Tr{\rho^{I}(\tau) \log_2[\rho^{I}(\tau)]}$, where $\rho^{I}(\tau) = \Tr_{\epsilon}\left\{\tilde \rho(\tau)\right\}$ is the $2\times 2$ reduced density matrix of the ion \cite{haroche_exploring_2013}.

\bibliography{paper_revised}

\end{document}